\documentclass[aps,prl,twocolumn,superscriptaddress]{revtex4-1}

\usepackage{graphicx}
\usepackage{amssymb,amsmath,wasysym}

\newcommand{\tn}{\textnormal}

\begin{document}

\title{A non-empirical free volume viscosity model for alkane lubricants under severe pressures}
\author{Kerstin Falk}
\affiliation{Fraunhofer IWM, W\"ohlerstr.\,11, 79108 Freiburg, Germany}
\author{Daniele Savio}
\affiliation{Fraunhofer IWM, W\"ohlerstr.\,11, 79108 Freiburg, Germany}
\author{Michael Moseler}
\affiliation{Fraunhofer IWM, W\"ohlerstr.\,11, 79108 Freiburg, Germany}
\affiliation{Institute of Physics, University of Freiburg, Herrmann-Herder-Str.\,3, 79104 Freiburg, Germany}

\begin{abstract}
Viscosities $\eta$ and diffusion coefficients $D_s$ of linear and branched alkanes at high pressures $P$$<$0.7 GPa and temperatures $T$=500-600 K are calculated by equilibrium molecular dynamics (EMD). Combining Stokes-Einstein, free volume and random walk concepts results in an accurate viscosity model $\eta(D_s(P,T))$ for the considered P and T.  All model parameters (hydrodynamic radius, random walk step size and attempt frequency) are defined as microscopic ensemble averages and extracted from EMD simulations rendering $\eta(D_s(P,T))$ a parameter-free predictor for lubrication simulations.
\end{abstract}

\maketitle

Knowlegde-based design and optimization of liquid lubricants require a quantitative modelling of their rheological properties under relevant tribological conditions~\cite{Bair2007}. 
For instance, lubricants in roller element bearing and gear applications are subject to pressures of the order of GPa~\cite{Spikes1994}. 
Traditional empirical viscosity models (such as Barus or Roelands equation) fail to describe $\eta(P)$ over the relevant pressure range~\cite{Bair2003} indicating that improved viscosity models for the extreme pressure regime \cite{Bair2013, DelaPorte2014} should be based on physical insights~\cite{Vergne2014}. 
A promising approach employs the Stokes-Einstein relation \cite{Einstein1905}
 \begin{equation}
   D_s = k_BT/(n\pi\eta R_h)
   \label{eq:SE}
 \end{equation}
that connects the viscosity with the self-diffusion coefficient $D_s$.
Here, $R_h$ denotes the hydrodynamic radius and $n$ lies between the slip and no-slip hydrodynamics limits, 4 and 6. 
The applicability of Eq.\,\eqref{eq:SE} on a microscopic level 
has been theoretically motivated and is well established under normal conditions~\cite{Zwanzig1970,Hansen2013}, 
also for non-spherical molecules if $R_h$ is considered a free parameter. 
However, a breakdown of Stokes law has been observed in various dense liquids, including molecular glass formers~\cite{Bordat2003,Brillo2011}. 
The mechanism of this breakdown is still subject to extensive research, mostly focused on densification obtained by supercooling~\cite{Charbonneau2014, Henritzi2015, Kawasaki2017}. 
This raises the question whether Eq.\,\eqref{eq:SE} remains valid for liquids which are densified by pressurisation instead of cooling. 

In this letter, EMD simulations are utilized to validate  Eq.\,\eqref{eq:SE} for linear and branched alkanes (constituents of ordinary lubricant base stock) for a pressure and temperature range representing typical tribological high load applications. 
We suggest a microscopic defintion of 
 \begin{equation}
   R_h = \sqrt{\langle a\rangle/\pi}
   \label{eq:RS}
 \end{equation}
by introducing an EMD averaged molecular cross section $\langle a\rangle$. This represents an important step towards a parameter-free structure-property relationship (SPR) relateing molecular structure with macroscale viscosities. 

Employing Eq. (1) necessitates an additional SPR for $D_s$. 
Here, we utilize the free volume (FV) concept~\cite{Doolittle1951,Cohen1959}
 \begin{equation}
   D_s = D_0 \exp(-v_c/v_f).
   \label{eq:FVT}
 \end{equation}
with the mean FV per molecule $v_f$ and the critical volume $v_c$. 
Note, that since $v_f$ is determined by the lubricant density $\rho$ an equation of state $\rho(P)$ is required to arrive at a pressure dependent viscosity law.  

Although widely used for soft matter systems~\cite{Almeida1992,Javanainen2010,Vrentas2013}, the FV concept is being challenged due to concerns about the relevance of FV compared to energetic effects and about the physical interpretation of the free parameters $D_0$ and $v_c$~\cite{Vrentas2003,Falck2005,Betancourt2015,Berthier2011,Jadhao2017}. 
We show in the second part of this letter that Eq.\,\eqref{eq:FVT} can be applied to our alkane lubricants and suggest microscopic definitions of $D_0$ and $v_c$.  
The latter relies on the observation that self-diffusion can be considered a random walk of a molecule's center of mass (COM) with step length and attempt frequency determined by EMD simulations.  

\begin{figure*}
  \begin{center}
    \includegraphics[width=\textwidth]{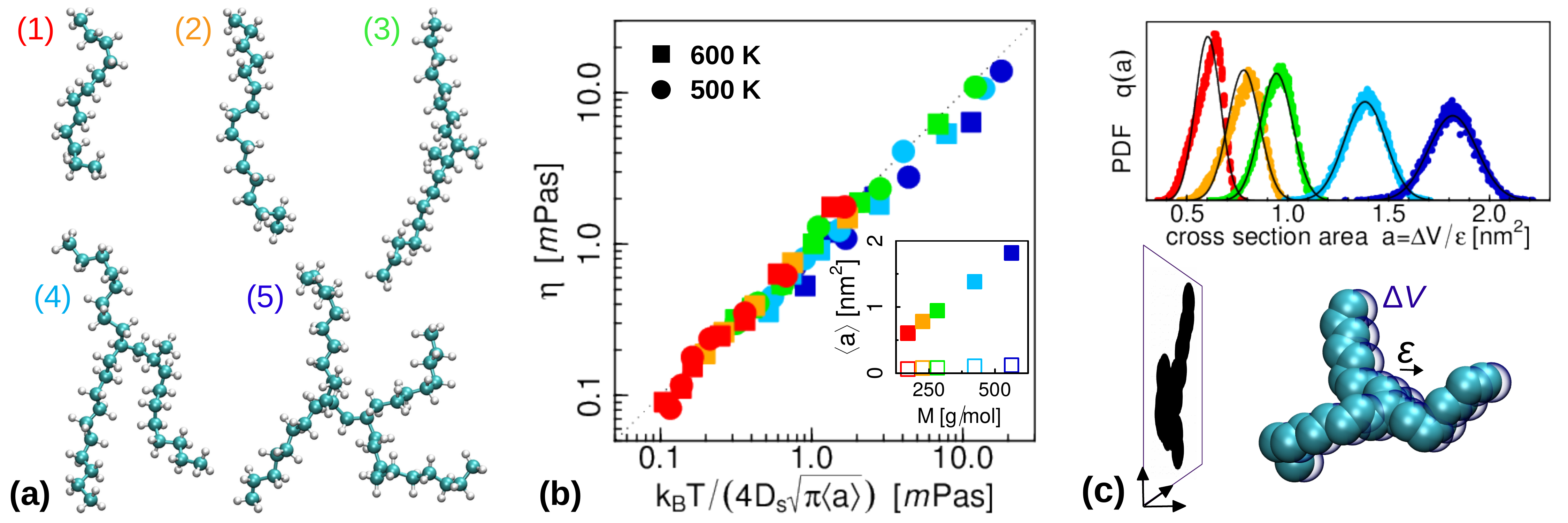}
    \caption{  \label{fig1}
	Stokes radius of complex molecules.
	{\it (a)} Considered molecular structures: 
	{\it n}-dodecane {\it (1, red)}, 
	{\it n}-hexadecane {\it (2, orange)}, 
	PAO-C$_{10}$-dimer {\it (3, green)}, 
	PAO-C$_{10}$-trimer {\it (4, light blue)}, and 
	PAO-C$_{10}$-tetramer {\it (5, dark blue)}  
	{\it (b)} Shear viscosity $\eta$ and self diffusion coefficient $D_s$ 
	under extreme pressure (up to $0.7$ GPa) and temperature ($\blacksquare/\CIRCLE$ $600/500$\,K) 
	from EMD simulations: 
	Quantitative agreement with the Stokes-Einstein relation Eq.\eqref{eq:SE}, 
	assuming a slip boundary condition ($n$=4) and a molecule radius $R_h=\sqrt{\langle a\rangle/\pi}$ 
	with $\langle a\rangle$ the molecule's mean cross section area
	({\it inset}: $\blacksquare$ mean $\langle a\rangle$, $\square$ standard deviation). 
	{\it (c)} Definition of the configuration dependent cross section area $a:=\Delta V / \varepsilon$ 
	($\Delta V$ newly occupied volume after a small virtual displacement $\varepsilon$), and resulting distributions $q(a)$; 
	for comparison, lines show Gaussian distributions.
    }
  \end{center}
\end{figure*}

Two linear alkanes and three poly-$\alpha$-olefines (PAO) (for structures see Fig.\ref{fig1}a) are modelled with the all-atom 
optimized potentials for liquid simulations (AA-OPLS)~\cite{Jorgensen1996}.
The EMD of these lubricants is simulated within a constant volume subject to periodic boundary conditions for densities ranging roughly from $470$ to $850$\,kg/m$^3$.
Time integration is performed employing the LAMMPS software suite~\cite{Plimpton1995, LAMMPS} 
with timestep $0.5$\,fs, and a Nos\'e-Hoover thermostat with relaxation time $0.1$\,ps~\cite{AllenTildesley1989,FrenkelSmit2001}.
Viscosities $\eta$ are determined via the Green-Kubo formalism and self-diffusion 
coefficients $D_s$ via the mean squared displacement (MSD)\cite{Hansen2013} (see supp.\,Figs.\,S1/2). 

As shown in Fig.\,\ref{fig1}b, results for $\eta$ and $D_s$ vary over 3 orders of magnitude and 
are fully compatible with Eq.\,\eqref{eq:SE} (assuming slip boundary conditions $n$=4~\cite{Hansen2013,Zwanzig1970}).
A parameter-free quantitative agreement is achieved by introducing the hydrodynamic radius via Eq.\,\eqref{eq:RS} as follows.
Since the Stokes drag on macroscopic solid objects with slippery surfaces scales with the object's cross section in the direction of a displacement~\cite{Leith1987}, we calculate  a directional molecular cross section $a$ as indicated in Fig.\,\ref{fig1}c.
To each molecule a volume $v_{mol}$ is assigned using a coarse grained hard sphere approach based on a CH$_X$ ($X=1,2,3$) united atom representation~\cite{Martin1998} (see details in supp.\,Fig.\,S3).
Then the molecule is displaced over a short distance $\varepsilon$ and its effective cross section area is defined 
by $a = \Delta V / \varepsilon$, where $\Delta V$ is the newly occupied volume.  
Finally, the mean cross section is obtained by $\langle a\rangle=\int a q(a) \tn{d}a$, where $q(a)$ is 
the probability for the molecule to have a configuration with cross section $a$. 
Interestingly, $\langle a\rangle$ scales linearly with the molecule size, despite the different morphologies (i.e. number of branches in alkanes - see inset in Fig.\,\ref{fig1}b). 
This scaling can be rationalized by a cylindrical shape estimate of long alkane chains, neglecting the contribution of chain ends and knots (supp.\,Fig.\,S4). 

\begin{figure}
  \begin{center}
    \includegraphics[width=0.5\textwidth]{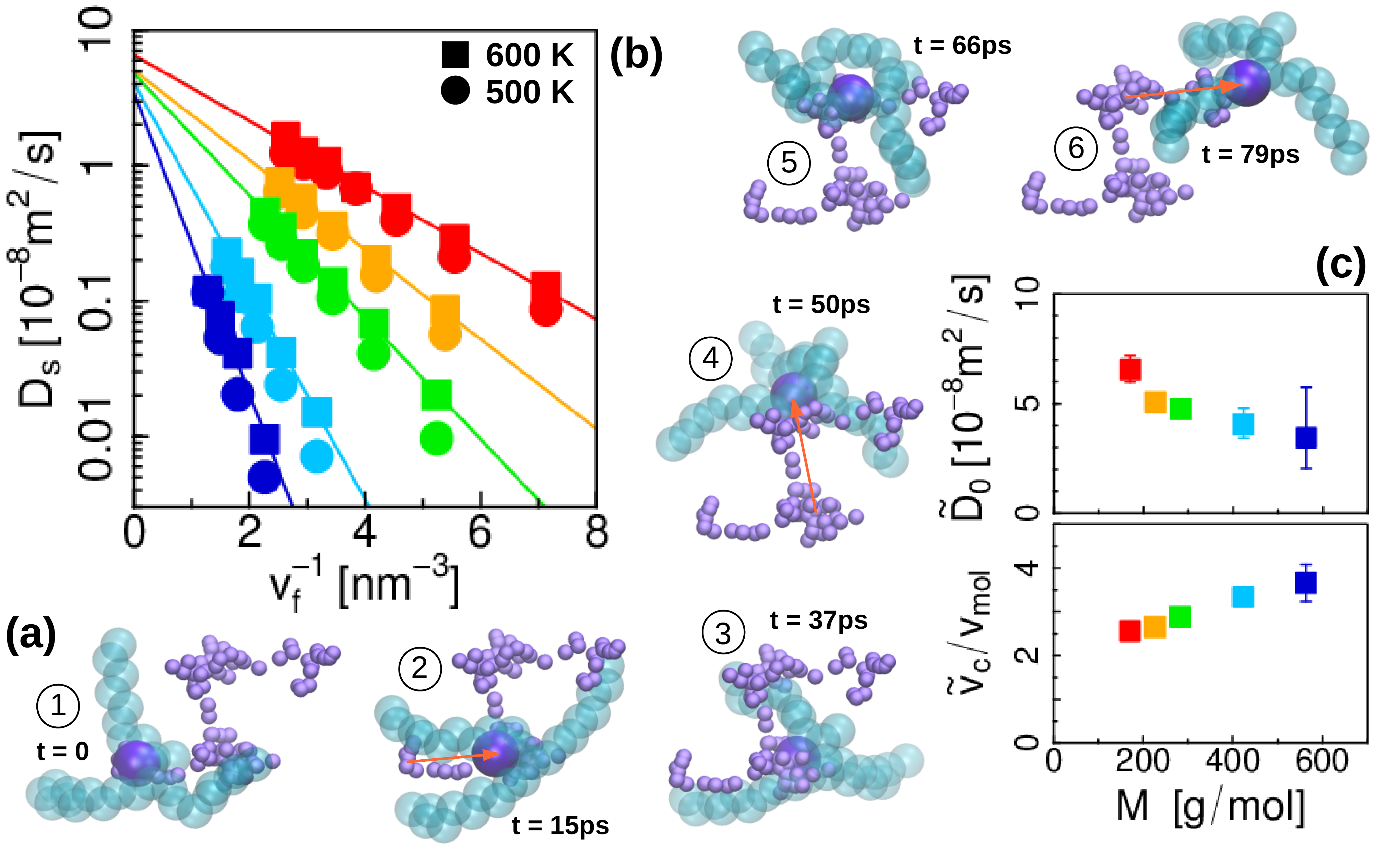}
    \caption{  \label{fig2}
	Dependence of diffusion on free volume. 
	{\it (a)} Part of a C$_{10}$-trimer trajectory within bulk fluid at $0.6$\,MPa. 
	Light blue: C-atoms; large dark blue: COM position; small violet: all COM positions (every $1$\,ps during $85$\,ps). 
	The COM diffusive motion (arrows) can be described by a random walk between caged positions due to the confining presence of the surrounding molecules 
	(not displayed).
	{\it (b)} Self diffusion coefficient $D_s$ {\it vs.} inverse of the mean free volume per molecule $v_f = v - v_{mol}$ 
	(colors as in Fig.\,\ref{fig1}, $\blacksquare/\CIRCLE$ $600/500$\,K).
	Lines are best fits of $D_s = D_0\exp(-v_c/v_f)$ on $600$\,K data. 
	{\it (c)} Fit results $\tilde{D_0}$ and $\tilde{v_c}$; errorbars: $68$\%-confidence interval. 
  }
  \end{center}
\end{figure}

After having established a parameter-free relation between $\eta$ and $D_s$, we now focus on the diffusive motion of the alkanes. 
Fig.\,\ref{fig2}a shows part of a C$_{10}$-trimer COM trajectory 
with a behaviour which is characteristic of a caging effect. 
The COM position oscillates within a compact volume due to confinement by the neighboring molecules (snapshots 1,3,5 of Fig.\,\ref{fig2}a). 
Elementary diffusion steps (EDS) take place via occasional irreversible translations (indicated by red arrows in snapshots 2,4,6 of Fig.\,\ref{fig2}a).
The FV ansatz leading to Eq.\,\eqref{eq:FVT} assumes that the probability 
for an EDS is given by 
$p(v_c) = \exp(-v_c/v_f)$.
Here, $v_c$ is the critical void size in the cage formed by a molecule's neighbours allowing for an irreversible COM jump. 
Following the simple argument that this void has to accommodate the molecule, the critical volume $v_c$ is expected to be of the order of the hard core molecule volume $v_{mol}$. 
Note, that $p(v_c)$ depends parametrically on the mean free volume per molecule  $v_f = v-v_{mol}$, where $v$ denotes the molar volume~\cite{Cohen1959}. 

Indeed, the self diffusion coefficient follows the form Eq.\,\eqref{eq:FVT}, 
as shown in Fig.\ref{fig2}b, where lines are best fits for the $600$\,K data. 
Fig.\,\ref{fig2}c displays 
the dependence of the fit parameters $\tilde{v}_c$ and $\tilde{D}_0$ on the size of the molecules. 
Surprisingly, the critical volume is about $3$ times larger than $v_{mol}$ in contradiction to the simple argument stated above.  
An alternative interpretation of $v_c$ is based on the following consideration. 
To perform an EDS, a molecule needs to move from its cage center to a void in the cage wall. 
The necessary critical volume for this displacement over the cage size $r_c$ is then $r_c\cdot a$ 
with $a$ the molecule's cross section. 
For constant $a$ and $r_c$, the diffusion process could then be pictured as a random walk with stepsize $r_c$ and step frequency $1/\Delta t$. 
The latter is the product of an attempt frequency $1/\tau_0$ and the success probability $\exp(-r_c a/v_f)$ resulting in
\begin{equation}
  D_s(a,r_c) = r_c^2/(6\Delta t) = r_c^2/(6\tau_0)\exp(-r_c a /v_f).
  \label{eq:Ds_ar}
\end{equation}
However, both $a$ and $r_c$ depend a priori on the molecules' configurations with respect to the direction of the EDS. 
Fluctuating shapes and distances in molecular fluids result in a probability distribution $q(a,r_c)$ for $a$ and $r_c$ that determine the diffusion coefficent $D_s(a,r_c)$ in a certain direction and thus the total diffusion coefficient is 
  $ D_s = \int D_s(a,r_c)q(a,r_c)\,\tn{d}a\,\tn{d}r_c$. 
In the following, we demonstrate by sampling $q(a,r_c)$ over all possible configurations and orientations that the first moments 
$\langle a\rangle=\int a q(a,r_c)\,\tn{d}a\,\tn{d}r_c$ 
and 
$\langle r_c\rangle=\int r_c q(a,r_c)\,\tn{d}a\,\tn{d}r_c$
dominate the diffusion process:
 \begin{equation}
   D_s  \approx \langle r_c\rangle^2/(6\tau_0)\exp (-\langle r_c\rangle\langle a\rangle/v_f)
  \label{eq:Ds_arave}.
 \end{equation}

\begin{figure}
  \begin{center}
  \includegraphics[width=0.5\textwidth]{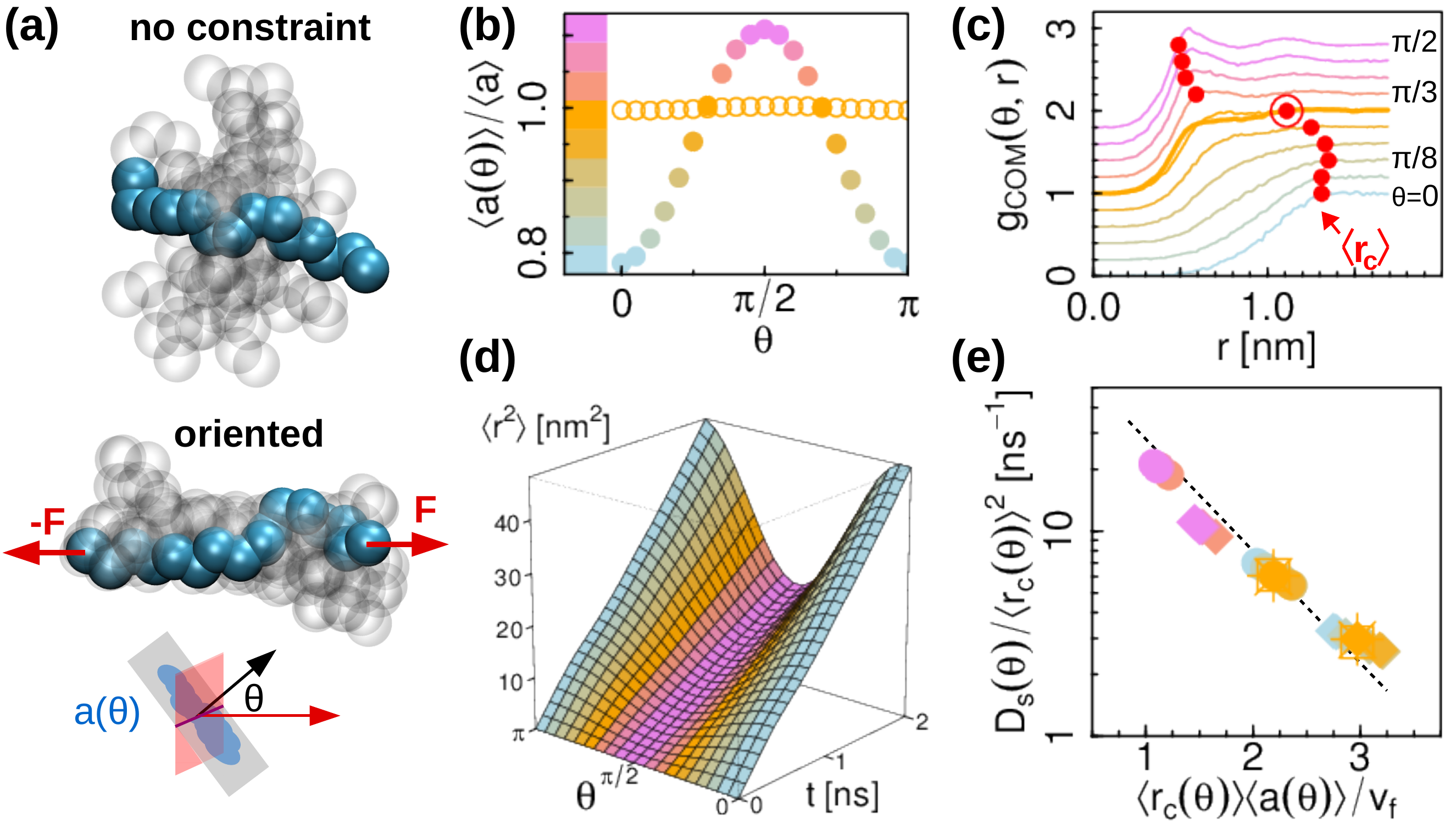}  
  \caption{  \label{fig3}
	Diffusion of artificially oriented {\it n}-hexadecane molecules in an unconstrained bath. 
	{\it (a)} Superimposed configurations of one randomly chosen bath and one oriented molecule ($\pm F$ applied along $\theta=0;\pi$, COM motion subtracted, $1$\,frame/ns). 
	{\it (b)} mean cross section $\langle a(\theta)\rangle$ for un-/constrained ($\Circle$/$\CIRCLE$) molecules, unperturbed value $\langle a \rangle$ see Fig.\ref{fig1}. 
	{\it (c)} $\theta$-dependent COM radial distribution function (shifted for better visibility) 
	with estimate of cage radius $\langle r_c(\theta)\rangle$ ($\bullet$); same for isotropic COM-RDF of bath molecules (bold line, $\Circle$).
	{\it (d)} Anisotropic MSD for oriented molecules in least dense system. 
	{\it (e)} Anisotropic self diffusion coefficient $D_s(\theta)$ (2 different densities $\CIRCLE$,$\blacklozenge$) 
	normalized with static structure properties according to Eq.\eqref{eq:Ds_arave}; 
	direction averaged values $D_s=(D_s(0)+2D_s(\pi/2))/3$, isotropic $D_s$ of bath molecules and of unperturbed systems ($\times,+,\Square$) are identical. 
  }
  \end{center}
\end{figure}

For a given configuration, Eq.\eqref{eq:Ds_ar} implies a direction dependent stepsize and success probability for an individual EDS. 
To study this anisotropy we consider an auxiliary system of preferentially oriented alkanes immersed in a bath of unconstrained molecules.
This artificial test situation is realized for {\it n}-hexadecane at two different densities by applying opposing external forces $\pm0.05$\,eV/\AA\ to the head and tail carbon atoms of $2.6\%$ 
of the  molecules (Fig.\ref{fig3}a). 
The resulting preferential orientation leads to a permanently anisotropic cross section $\langle a(\theta)\rangle$ (Fig.\ref{fig3}b) and cage radius $\langle r_c(\theta) \rangle$ (red dots in Fig.\ref{fig3}c), where $\theta$ denotes the angle between the applied forces and the EDS direction.
Here, $\langle a(\theta)\rangle$ was calculated as previously defined (see Fig.\,\ref{fig1}c) for a given direction $\theta$.  
Lacking an unambiguous definition of $\langle r_c(\theta) \rangle$, a pragmatic estimate was based on the direction dependent 
radial distribution function $g_{COM}(\theta,r)$ of the molecules' COM (Fig.\,\ref{fig3}c) via $g_{COM}(\theta,\langle r_c(\theta)\rangle)=1$. 
In the same spirit, the isotropic cage size $\langle r_c \rangle$ (circle in Fig.\,\ref{fig3}c) was determined from the isotropic 
radial distribution function $g_{COM}(r)$ (bold line in Fig.\,\ref{fig3}c). 
Note, that this value is close to the result obtained via  $\langle r_c \rangle=\int \langle r_c(\theta) \rangle d\,\cos{\theta}$.  

As expected the preferentially oriented molecules exhibit a pronounced anisotropy in the MSD $\langle r^2(\theta)\rangle$ (Fig.\ref{fig3}d) 
and consequently in the diffusion coefficient $D_s(\theta)$.
Interestingly, the mean diffusion coefficient $D_s=(D_s(\theta=0)+2D_s(\theta=\pi/2))/3$  
is equal to the isotropic diffusion coefficient of the unperturbed molecules, confirming that diffusion is given by an average over all directions with respect to the main axis of a hexadecane molecule. 
Most importantly however, $D_s(\theta)$ scales with the respective static structure properties as predicted by Eq.\eqref{eq:Ds_arave}  validating the applicability of a FV ansatz on the microscopic level of an EDS (Fig.\,\ref{fig3}e).  

\begin{figure*}
  \begin{center}
  \includegraphics[width=\textwidth]{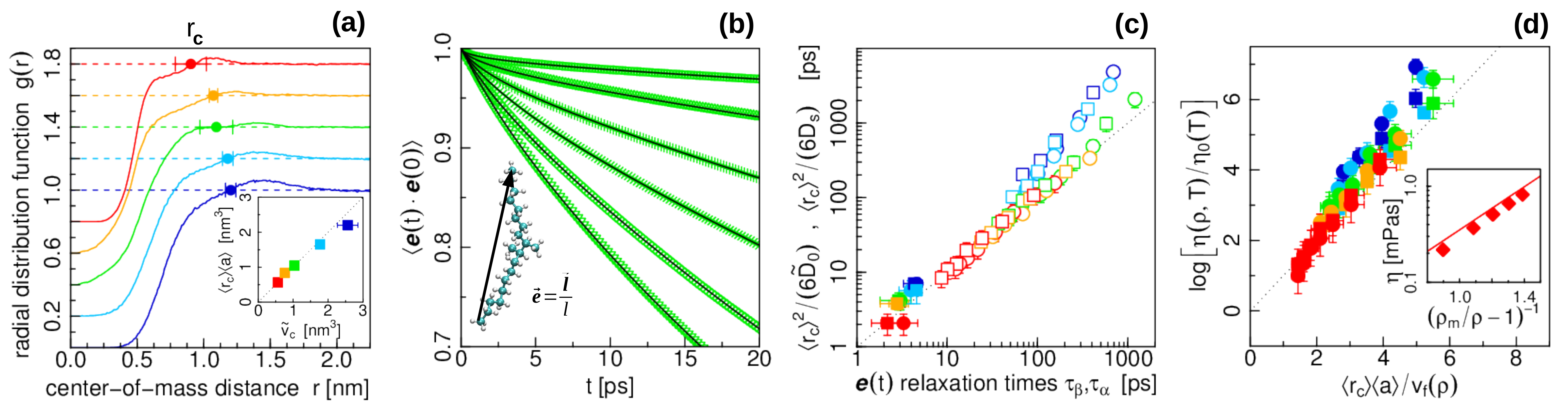}  
  \caption{  \label{fig4}
	Scaling of viscosity and self diffusion coefficient with molecule structure properties 
	[Colors as in Fig.\,\ref{fig1}].
	{\it (a)} COM radial distribution functions (lines, shifted for better visibility) with 
	estimate of $\langle r_c\rangle$ ($\bullet$);
	{\it inset:} fit result for the critical volume $\tilde{v}_c$ (see Fig.\,\ref{fig2}) 
	{\it vs.} $\langle r_c\rangle\langle a\rangle$ ($\langle a\rangle$ mean cross section area, see Fig.\,\ref{fig1}).
	{\it (b)} Time autocorrelation function of the C$_{10}$-dimer end-to-end 
	vector orientation ${\bf e}(t)$ for all considered densities at $500$\,K (symbols) 
	with fits of Eq.\,\eqref{eq:eteacf} (lines); all other molecules in supp.\,Fig.\,S7. 
	{\it (c)} Correlation times $\tau_\beta$ and $\tau_\alpha$ (full/empty symbols) 
	{\it vs.} predicted attempt and waiting time of the random walk diffusion model 
	$\tau_0 = \langle r_c\rangle^2/(6D_0)$ and $\Delta t = \langle r_c\rangle^2/(6D_s)$, respectively; 
	data for $\tau_\beta$ shows the mean value from all densities. 
	{\it (d)} Scaled viscosity $\log(\eta/\eta_0)$ with 
	$\eta_0 = 1.5k_BT\tau_\beta/(\langle r_c\rangle^2 \sqrt{\pi\langle a\rangle})$ {\it vs.} 
	ratio of critical to free volume $\langle r_c\rangle \langle a\rangle /v_f$ (no free parameter); 
	{\it inset:} Experimental data for {\it n}-dodecane at $T=473$\,K from Ref.\cite{Caudwell2004} ($\blacklozenge$) and prediction from simulations (line).
  }
  \end{center}
\end{figure*}

Comparing Eq.\eqref{eq:FVT} with \eqref{eq:Ds_arave} leads to $D_0=\langle r_c\rangle^ 2/(6\tau_0)$ and $v_c=\langle r_c\rangle\langle a\rangle$.  
Indeed, applying the above structure evaluation for $\langle a\rangle$ (inset in Fig.\,\ref{fig1}b) and $\langle r_c\rangle$ (Fig.\,\ref{fig4}a) to the unperturbed systems of all 5 fluid types reveals that the product $v_c=\langle r_c\rangle\langle a\rangle$ 
agrees well with the fitted critical volumes $\tilde{v}_c$ (Fig.\,\ref{fig4}a inset). 
Both, $\langle a\rangle$ and $\langle r_c\rangle$ are only weakly dependent on density and temperature (supp.\,Figs.\,S5/6) and can be conveniently estimated from an EMD simulation for a single $\rho$ and $T$.   

The remaining free parameter $\tau_0$ (time between random walk attempts) can be interpreted as the time 
scale for structure decorrelation in the molecule/cage system.
The connection between structural relaxation, diffusion and viscosity is subject of ongoing research~\cite{Ma2019}, and fully unravelling the underlying mechanisms 
goes far beyond the scope of this work. 
Nevertheless, as a starting point we consider the time autocorrelation function $\langle {\bf e}(0) \cdot {\bf e}(t)\rangle$
of the molecules' end-to-end vector orientation ${\bf e}(t)$ (see Fig.\,\ref{fig4}b) 
to quantify the intramolecular structure decorrelation.  
For long hydrocarbon chains, $\langle {\bf e}(0) \cdot {\bf e}(t)\rangle$ is well described 
by a double exponential function~\cite{Morhenn2012}
 \begin{equation}
  \langle {\bf e}(0) \cdot {\bf e}(t)\rangle = C \tn{e}^{-t/\tau_\alpha} + (1-C)\tn{e}^{-(t/\tau_\beta)^b}
  \label{eq:eteacf}
 \end{equation}
with two separate characteristic decay times $\tau_\alpha$ and $\tau_\beta$. 
On the one hand, a long time decay is observed on the time scale of the diffusion process 
$\tau_\alpha\approx \Delta t = r_c^2 / (6 D_s)$, ranging from $10$\,ps - $1$\,ns. 
On the other hand, $\tau_\beta$ is of the order of $1-10$\,ps and is insensitive to the fluid density 
(within statistical uncertainties, see supp.\,Fig.\,S7). 
This $\beta$-relaxation time fits well with the expected attempt frequency for 
the random walk $\tau_\beta \approx \tau_0 = r_c^2 / (6 D_0)$ as illustrated in Fig.\,\ref{fig4}c, 
which suggests it as a good measure for the relevant structure decorrelation on short times. 

The presented results are further validated in a series of scaling tests with modified  
model parameters for both intra- and intermolecular interactions (supp.\,Fig.\,S8). 
In particular, the strength of nonbonded interactions has little influence, 
but a scaling of the atomic radii $\sigma_{LJ}$ results in strong variations of the diffusivity
caused by the exponential term in Eq.\,\eqref{eq:Ds_arave}. 
Moreover, the prefactor $D_0$ is sensitive to variations of the energy barrier for bond rotation, 
which influences intramolecular relaxations. 
All scaling tests are also in quantitative agreement with the Stokes-Einstein 
relation, supporting our definition of a hydrodynamic radius in Eq.\,\eqref{eq:RS}.

Finally, combining Eqs.\eqref{eq:SE}, \eqref{eq:RS} and \eqref{eq:Ds_arave} 
the viscosity can be expressed as a function of density 
 \begin{equation}
  \log \frac{\eta(T,\rho)}{\eta_0(T)}= \frac{\langle r_c\rangle \langle a\rangle}{v_{mol}}(\rho_m/\rho-1)^{-1}
  \label{eq:eta rho}
 \end{equation}
with $\rho_m = M/v_{mol}$ the maximum hypothetical density for zero free volume 
($1-v_f/v = \rho/\rho_m$; supp.\,Fig.\,S4). Apart from the density, 
the r.h.s. contains only equilibrium structure properties, namely the
molecules' volume $v_{mol}$, mean cross section $\langle a\rangle$ and mean next neighbor distance $\langle r_c\rangle$. The temperature dependence of  $\eta$ enters via 
$  \eta_0(T) = 1.5 k_BT \tau_0(T)/(r_c^2 \sqrt{\pi \langle a\rangle})$.
By employing Eq.\eqref{eq:eta rho} and identifying $\tau_0$ with $\tau_\beta$ a parameter free rescaling of the simulated viscosities 
can be established (Fig.\,\ref{fig4}d).
This scaling law can also be applied to experimental high $T$ and $P$ viscosity data for {\it n-}dodecane~\cite{Caudwell2004}.
We find a good agreement of the experimental data with our parameter-free viscosity model Eq. \eqref{eq:eta rho} (inset of Fig.\,\ref{fig4}d).

To conclude, a combination of basic random walk and FV theory fully describes the self diffusion mechanism of long alkane chains, 
linear and branched alike, in the high $T$ and $P$ regime. 
A crucial part of the presented work is the introduction of the mean cross section $\langle a \rangle$ and mean cage size $\langle r_c \rangle$ as 
novel molecule shape parameters. 
While $\langle a \rangle$ establishes a quantitative link between viscosity and 
self diffusion via the Stokes-Einstein relation,  $\langle a \rangle$ and $\langle r_c \rangle$ allow for a parameter-free density scaling 
of both transport coefficients. 
The viscosity model can be directly implemented in density-based Reynolds-solvers~\cite{Elrod1981} 
and will contribute to a new cutting-edge simulation tool for tribological applications. 
The proposed shape parameters also opens new possibilities to quantify the role of molecular 
structure on rheology in anisotropic situations, such as shear thinning~\cite{Liu2017,Ingebrigtsen2018} 
or in nanometer-thin boundary lubrication films~\cite{Rosenhek-Goldian2015}.
Our approach might also be useful for other soft materials, such as self assembled membranes~\cite{Almeida1992,Javanainen2010}, 
polymer--solvent systems~\cite{Vrentas2013} or adsorbates in nanopores~\cite{Falk2015}. 

The authors gratefully acknowledge funding by the industrial partners of the MikroTribologie Centrum $\mu$TC (Karlsruhe, Germany), 
computing time within project HFR14 at NIC J\"ulich and  
useful discussions with L.\,Joly, S.\,Kapfer and L.\,Bocquet. 

\bibliographystyle{apsrev4-1}
\bibliography{References}

\end{document}